\documentclass[runningheads]{llncs}
\usepackage[T1]{fontenc}
\usepackage{rotating}
\usepackage{color}
\usepackage[]{lipsum}
\usepackage{pdfpages}
\usepackage[]{subcaption}
\usepackage{bookmark}
\usepackage{bm}
\usepackage{multirow}
\usepackage{hyperref}

\usepackage[ruled,vlined]{algorithm2e}
\usepackage{dsfont}
\usepackage{amsmath}
\usepackage{amsfonts}
\usepackage{mathtools}
\usepackage{tabularx}
\usepackage{wrapfig}
\usepackage[capitalise]{cleveref}
\newcommand{\second}[1]{{\color{red}{#1}}}
\newcommand{\best}[1]{{\color{blue}{#1}}}

\begin{document}
%
\title{Hybrid training of denoising networks to improve the texture acutance of digital cameras }
%
\titlerunning{Hybrid training of denoising networks for camera evaluation}
%
\author{Raphaël Achddou\inst{1} \and Yann Gousseau\inst{1} \and Saïd Ladjal\inst{1}}
\institute{LTCI, Telecom Paris, Institut Polytechnique de Paris}

\authorrunning{Achddou et al.}
%

%
\maketitle              
\begin{abstract}
In order to evaluate the capacity of a camera to render textures properly, the standard practice, used by classical scoring protocols, is to compute the  frequential response to a dead leaves image target, from which is built a {\it texture acutance metric}. In this work, we propose a mixed training procedure for image restoration neural networks, relying on both natural and synthetic images, that yields a strong improvement of this acutance metric without impairing fidelity terms.    
The feasibility of the approach is demonstrated both on the denoising of RGB images and the full development of RAW images, opening the path to a systematic improvement of the texture acutance of real imaging devices.  

\keywords{Image denoising  \and Deep learning \and Image quality assessment}
\end{abstract}
\section{Introduction}


In order to correctly visualize a photograph, its corresponding RAW image undergoes a complex sequence of development operations including white balancing, demosaicking, tone mapping, and image restoration operations such as deblurring and denoising.
Camera manufacturers implement proprietary algorithms fine-tuned for each setting of each camera. 
As a result, the overall image quality is a combination of hardware characteristics (quality of the lens, size of the sensor) and software performances.
In order to fairly assess the quality of an imaging device, independent agencies have defined standard tests and ISO protocols. Each of these tests focus on a specific characteristic such as chromatic aberrations, noise reduction, or texture rendering.

Recently, with the increase in computational power and the advent of deep learning for image processing, more and more digital image processing stages can be replaced by learned neural networks \cite{gharbi2016deep}.
Recent works already aim at completely replacing the full image development pipeline with a single neural network, producing impressive results in standard conditions \cite{ignatov2020replacing} or extremely low-light conditions \cite{chen2018learning}.
Moreover, light neural network architectures can now be integrated in embedded systems, e.g. on smartphone devices. Neural methods present another key advantage: one can easily optimize their response to specific test images, by including them in training databases. 

For the specific task of texture rendering evaluation, Cao et al. \cite{cao2009measuring} first presented a protocol quantifying the ability of an imaging pipeline to preserve texture information. This is obtained through the frequential response of the system to dead leaves images with a specific perceptual metric called \emph{texture acutance}.
These images are known for their invariance properties, as well as statistical properties making them close to natural images (non Gaussianity, scaling property, distribution of the spectrum and gradient), as studied in \cite{alvarez1999size,gousseau2007modeling,lee2001occlusion}. 
This quality evaluation protocol later became an ISO standard to measure the preservation of textures\cite{ISO} and is now used by classical camera scoring protocols. In a different direction, Achddou et al. \cite{achddou2021synthetic} showed that image restoration networks could be trained from synthetic images only, using databases of dead leaves images. 

Inspired by these results, we propose, in this paper, to train a denoising neural network on natural and dead leaves images, to jointly optimize a new metric derived from the texture acutance and the classic data fidelity metrics on natural images. 
After presenting some related works on image restoration in \cref{sec:related_works}, we  first introduce in \cref{sec:acutance} the texture acutance metric and the corresponding perceptual loss for image restoration networks.
We then show in \cref{sec:denoising_acutance} that we can strongly improve the texture acutance metric without impairing performances on natural images, first for the task of Additive White Gaussian Noise removal (AWGN) and then for the development of RAW images. These results open the path to an automatic improvement of standard quality evaluation tests.

\section{Related Works}\label{sec:related_works}

The goal of image restoration is to retrieve a clean image from distorted observations. 
In many cases, the distortion process can be modeled as follows:
$y = Ax + n,$ where $x$ is the theoretically perfect image, $y$ the distorted observation, $A$ is a linear operator and $n$ is some noise.

In order to solve this problem, a first class of methods are based on prior hypotheses on the distribution of natural images. 
These methods try to impose regularity properties on the restored solutions.
For instance, wavelet shrinkage methods \cite{DavidL.Donoho1994,donoho1998minimax} or DCT-filtering methods \cite{Yu2011} reconstruct an image assuming that the targeted images can be well approximated by a sparse decomposition.
In turn, variational methods based on the total variation \cite{rudin1992nonlinear,chambolle2004algorithm} assume that the image gradient follows a Laplacian distribution.
Based on the assumption of self-similarity, non-local methods leverage the redundancy in the image content. This is either done by weighted averaging (Non Local Means \cite{buades2005non} Non Local Bayes \cite{lebrun2013nonlocal}) or by collaborative filtering (BM3D \cite{Dabov2007a}).

Over the past decade, learning-based approaches for image restoration have developed drastically.
After the success of neural networks for high-level computer vision tasks \cite{he2016deep}, these methods have been adapted to image restoration through the use of generative models \cite{zhang2017beyond,zhang2018ffdnet}.
Rather than using prior hypotheses, the parameters of the neural networks are tuned in a long optimization process to directly minimize the reconstruction error in a black box manner.
For the training to succeed, these methods require large databases of pairs of distorted and clean images.
Even though they are hard to interpret, they surpassed prior-based methods on most image restoration benchmarks by a large margin for a wide variety of tasks such as image denoising \cite{zhang2018ffdnet}, demosaicking \cite{gharbi2016deep} etc.

Following these initial works, recent papers extended the use of deep learning methods to real-world problems of image restoration such as RAW image denoising \cite{liu2021invertible,anwar2019real}.
Ignatov et al. \cite{ignatov2020replacing} and Chen et al. \cite{chen2018learning} also propose to fully replace the image development pipeline by a learned neural network, producing surprisingly good results.
However, acquiring  datasets of real-world pairs of distorted and clean RAW images is a cumbersome task \cite{abdelhamed2018high,chen2018learning}, which often requires complex post-processing algorithms. In order to ease the training process, a more restrained approach consists in modeling the distortion process accurately, and to synthesize them accordingly \cite{wei2021physics}. 

Going further, Achddou et al. \cite{achddou2021synthetic,achddou_cviu} proposed to train image restoration neural networks on generated dead leaves images  in order to completely circumvent the data acquisition process, reaching performances close to the networks trained on real images, for various image restoration tasks.
These images indeed exhibit statistical properties close to those of natural images \cite{alvarez1999size,gousseau2007modeling,lee2001occlusion} even though they depend from few parameters.
Following \cite{achddou2021synthetic}, similar synthetic databases were also used to pre-train image classification networks \cite{baradad2021learning} and disparity map estimators \cite{madhusudana2021revisiting}.
Prior to these works, dead leaves images were used to assess the capacity of cameras to render textures properly. This idea was first presented in 2009 by Cao et al. \cite{cao2009measuring}, which was later improved in the following references \cite{cao2010dead,artmann2015image}. We will present in detail these works in the following section.


\section{Texture acutance : a frequential loss assessing texture preservation}\label{sec:acutance}

\subsection{Dead leaves images}
\begin{figure}[htp]
    \centering
    \includegraphics[width = 0.75\textwidth]{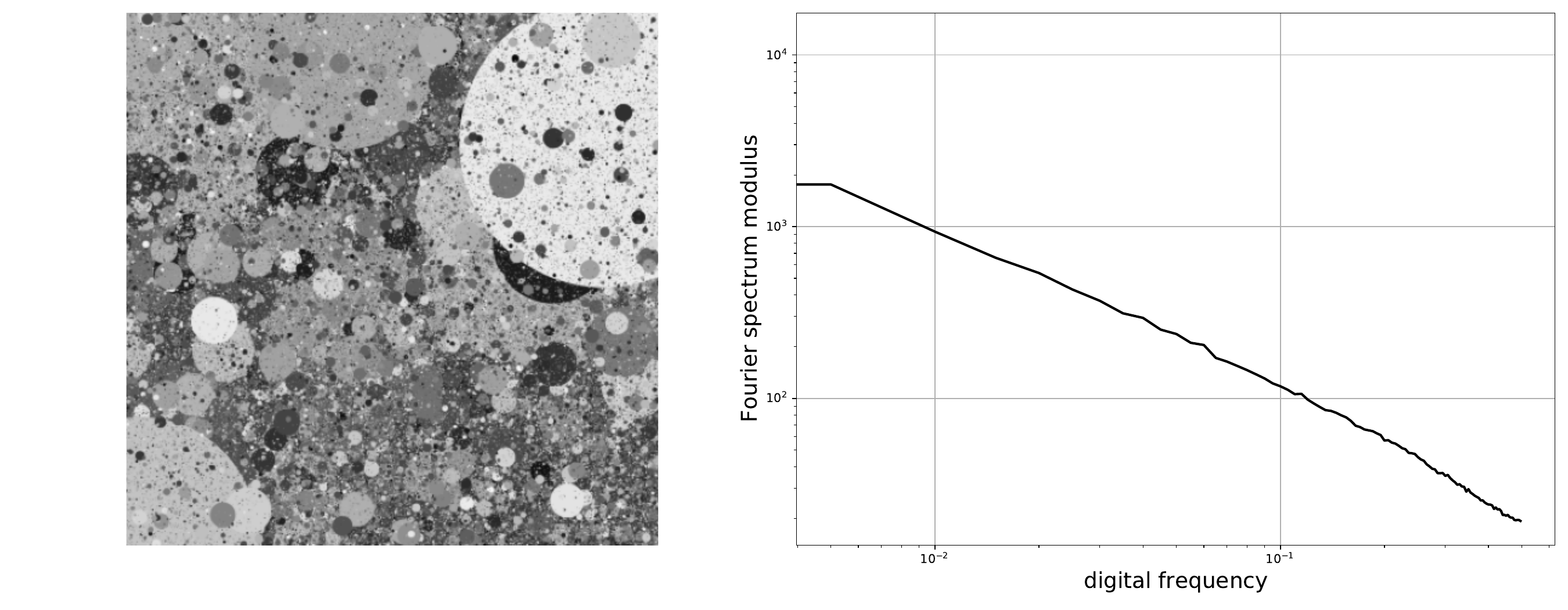}
    \caption{\footnotesize Grey level dead leaves image and its associated digital spectrum in logarthmic scales.
    The theoretical value is a straight line.}
    \label{fig:dead_leaves_spectrum}
\end{figure}

Dead leaves images were first introduced by Matheron in 1975 \cite{matheron75}, with the aim of modeling porous media. It was later shown that if object sizes fulfill some scaling property, this model accounts for many statistics of natural images~\cite{alvarez1999size,lee2001occlusion}.
To generate such images, shapes of random size, color and position are superimposed on top of each other until the whole image plan is covered.
In the simplest set-up, these shapes are disks of random radiuses. An example of a dead leaves image is given in \cref{fig:dead_leaves_spectrum}, along with its spectrum. A precise mathematical formulation of dead leaves images is given in \cite{bordenave2006dead}. 

Dead leaves images were first used for camera evaluation in 2009 by Cao et al \cite{cao2009measuring}.
The proposed idea is to measure the response of a camera to a specific image target.
Because of their invariances and statistical properties, the dead leaves model was chosen by the authors as the generation algorithm for the target.
Among the desired properties, scale invariance is achieved when the disks radii follows a power law with $\alpha = 3$. The dead leaves target is therefore generated with this parameter.
Note that to ensure the convergence of the algorithm, bounding parameters $r_{min},r_{max}$ are required \cite{gousseau2007modeling}.

\subsection{Texture acutance}

\begin{figure}[htp]
    \centering
    \includegraphics[width = 0.8\textwidth]{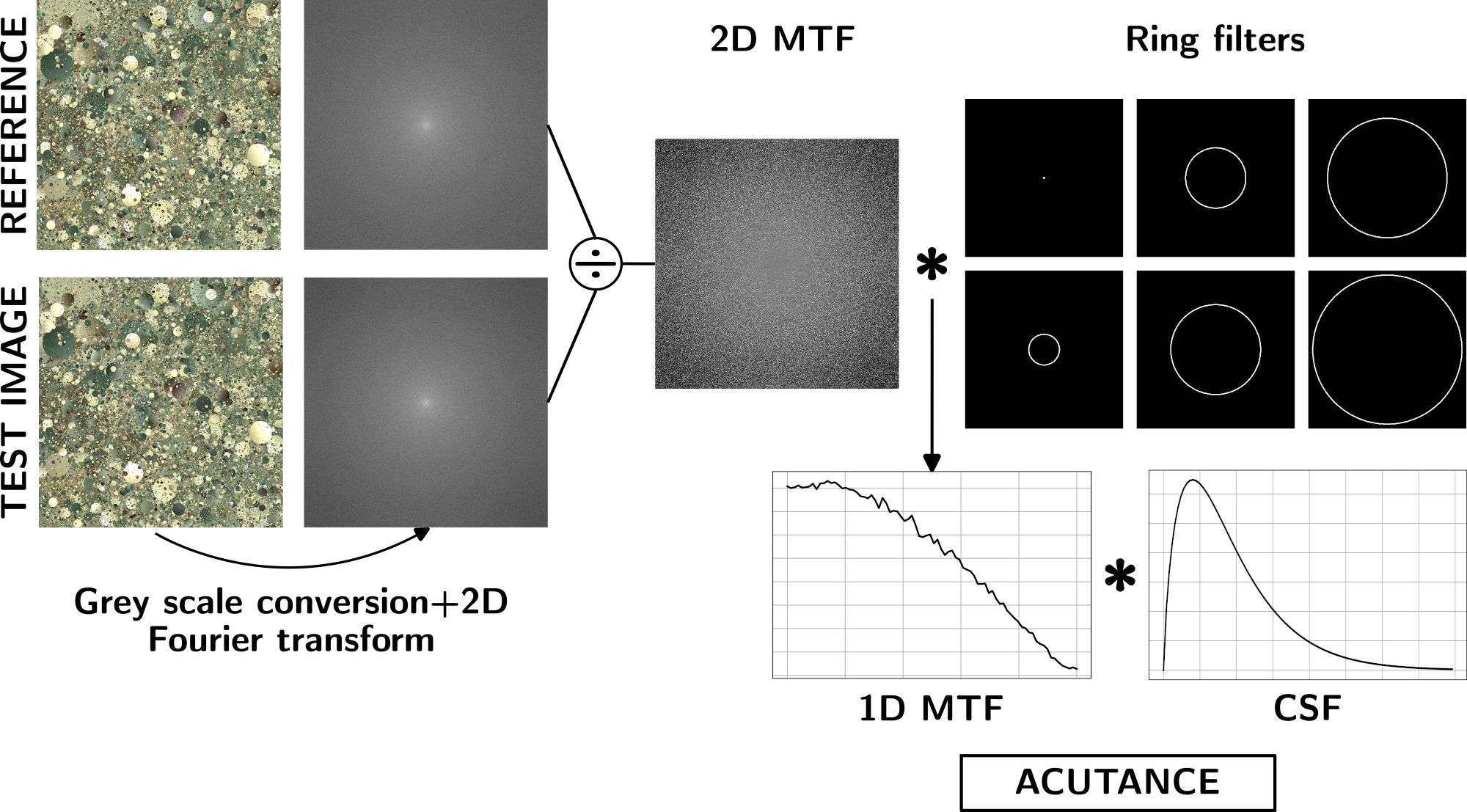}
    \caption{\footnotesize Diagram explaining the computation of the acutance metric}
    \label{fig:acutance_loss_diagram}
\end{figure}

In \cite{cao2009measuring}, the authors evaluate the response of a camera to the dead leaves target by computing the ratio of the power spectra, resulting in a Modulation Transfer Function (MTF).
At each position $(m,n)$ for an $(N,N)$ image:
$$\text{MTF}_{2D}(m,n) = \frac{|\hat{Y}(m,n)|}{|\hat{X}(m,n)|},$$ where $\hat{Y}$ is the spectrum of the obtained image and $\hat{X}$ is the ground truth spectrum.
In all that follows, we compute the image spectra on a greyscale version of the color image, obtained by the standard linear combination $\text{Grey} = 0.2126 R + 0.7152 G + 0.0722 B$.
The classical idea behind the MTF is that the ratio of the power spectra corresponds to the Fourier coefficients of the blur kernel induced by the camera, excluding non linear transforms often involved in the image development, as well as the impact of noise.

In order to account for the impact of noise in the estimation of the MTF, Artmann first proposed a corrected version of the MTF, by subtracting an estimate of the noise spectrum. However, the latter was computed by taking a photograph of a uniform grey surface, assuming an additive and signal independent noise model, which is far from reality. Moreover, some image development pipelines include a nonlinear noise reduction operation, which affect the estimation of the real MTF.

In the same paper \cite{artmann2015image}, Artmann proposes a new computation trying to correct these issues. 
Here, we consider the complex spectrum of a reference digital dead leaves target $\hat{X}$, rather than the estimate of the power spectrum $|\hat{X}|$ in the spatial domain. 
In the previous version, the phase information was lost. This is necessary in the context of camera calibration since phase information is reliable only if a registration algorithm is applied. In the context of training a denoising algorithm registration between noisy and restored image is supposed perfect and dealt with by the MSE-loss. 
Relying only on the amplitude of the spectrum meant that we could not differentiate frequencies which were already in the target and information that was added by the imaging device. Therefore noise and non linear functions had an impact in previous computations.

The proposed method, which we call $\text{MTF}_{cross}$ uses the cross power density between the target and the obtained image $\phi_{XY}(m,n)$, and the auto power density $\phi_{XX}(m,n)$. 
More precisely, 
$$ \phi_{XY}(m,n) = \hat{Y}(m,n)\hat{X}^*(m,n) \quad\text{and}\quad \phi_{XX}(m,n) = \hat{X}(m,n)\hat{X}^*(m,n).$$
Given these quantities, the MTF becomes :
\begin{equation}
\label{eq:mtfcross}
\text{MTF}_{cross}(m,n) = \left|\frac{\phi_{XY}(m,n)}{\phi_{XX}(m,n)}\right|.
\end{equation}


Since the dead leaves target is rotationally invariant, so is its spectrum. We therefore express the MTF as a 1D function by averaging it on concentric rings of width 1.
The MTF becomes :
$$\text{MTF}_{1D}(k) = \frac{\mathds{1}_{C_k}}{\#C_k} \times  \text{MTF}_{cross},$$ 
where $C_k = \bigl\{ (i,j) \in [-N/2,N/2]^2 | (k-1)^2 \leq |i^2+j^2| < k^2 \bigr\}$ corresponds to a ring of radius $k$ and $\# C_k$ is its cardinal.


Though the full $\text{MTF}_{1D}$ is a good indicator of the camera's capacity to render textures, it is more helpful to compute a single score.
To that end, the \emph{texture acutance} \cite{cao2009measuring} is defined as a weighted sum of the $\text{MTF}_{1D}$, with weights defined by a contrast sensitivity function (CSF), inspired by the slanted edge Spatial Frequency Response (SFR), used to evaluate the sharpness of a camera.
Our visual system is indeed more sensitive to some frequencies than others. 
In that regard, the CSF models the sensitivity of the visual system to spatial frequencies expressed in cycle/degree. 

Based on the physiological analysis of the contrast sensitivity of infants and monkeys led by Movshon and Kiorpes \cite{movshon1988analysis}, the chosen formula to model the CSF is :

$\text{CSF}(\nu) = a.\nu^{c}.e^{-b\nu} ,$ 
where $\nu$ is a spatial frequency expressed in cylces/degree, parameters are fixed as $b = 0.2$, $c = 0.8$, and $a$ is a normalizing parameter so that $\int_0^{\text{Nyquist}} \text{CSF}(\nu)d\nu = 1,$ where $\text{Nyquist}$ corresponds to the theoretical maximum spatial frequency of the device.
Given this formula, the texture acutance score can be written as :
$$ A = \int_0^{\text{Nyquist}} \text{CSF}(\nu).\text{MTF}_{1D}(\nu) d\nu.$$

Note that we need to convert spatial frequencies in cycles/degree to a digital frequency in cycles/pixel for homogeneity. To do so we use the following formula: $f_{spatial} = \frac{1}{\alpha} f_{digital}$, where $\alpha$ is the viewing angle. The latter depends on viewing conditions with the equality $\alpha = \frac{180}{\pi}\text{arctan}(\frac{P}{D})$, where $P$ is the pixel size and $D$ is the viewing distance, assumed to be equal to 0.2 mm and 1m respectively. This corresponds to a maximal spatial frequency of 40 cycles/degree which is approximately the limit of the human visual system.

The perfect MTF corresponds to a constant function equal to 1, meaning that the frequential content has been perfectly restored by the camera for every frequency. This leads to an acutance $A = 1$. 
An acutance greater than 1 indicates that some frequential content was added to the image, probably because of noise or sharpening. 
An acutance lower than 1 indicates that some frequencies have been lost.

\subsection{Acutance loss for image restoration CNNs}
In \cite{achddou2021synthetic}, the authors showed that models trained on mixed databases (natural and synthetic images) perform on par with models trained on natural images only, while improving results on dead leaves image targets. We believe we can improve the frequential response of models trained on mixed sets, by using the acutance score in a loss function.

In the context of AWGN removal for color RGB images, the noisy image corresponds to $Y = X+n$ where X is a ground truth dead leaves image of size $(N,N,3)$. 
The denoising network $f_{\theta}$ produces an estimate of the clean image $Z = f_{\theta}(Y)$. For our restoration problem, we can consider that the denoising network is analogous to the camera which acquires the dead leaves target. 
We can compute  $\text{MTF}_{cross}$
for the denoising network using Formula~\eqref{eq:mtfcross}, based on the computation of the digital spectrum of both $X$ and $Z$. 

The obtained $\text{MTF}_{cross}$ is turned into a 1D signal as described above. For faster computation, concentric ring masks are stored in GPU so that the computation of $\text{MTF}_{1D}$ can be accelerated with parallel computing. 
Since the best possible acutance is 1, we define the acutance loss function as :
$$ \mathcal{L}_{acutance}(Y,X) = | 1 - A(f_{\theta}(Y),X)|,$$
which penalizes both adding or removing frequential information. In order to get a complete loss function, we add to it the $\mathcal{L}_2$ loss (i.e. the Euclidian distance), the initial fidelity term of the network. Indeed, the acutance loss $\mathcal{L}_{acut}$ is computed solely on an aggregation of the Fourier spectrum and is therefore blind to the spatial organisation of the image and can not replace an MSE-loss. 
When training on dead leaves images the  loss is therefore  
$$\mathcal{L} = \mathcal{L}_2 + \lambda.\mathcal{L}_{acut},$$
where $\lambda$ is a weighting parameter.

Since we train the image denoiser on both natural images and dead leaves images, we compute the acutance loss only on the dead leaves images in a minibatch $D$ of size $K$ and the $\mathcal{L}_2$ loss for all images. The formation of minibatches during training indeed randomly samples images from the mixed set. Thus, the loss in a batch becomes :
\begin{equation}
\label{eq:lossbatch}
\mathcal{L}_{batch} = \frac{1}{K}\sum_{i = 0}^{K} ||x_i - f_{\theta}(x_i+n_i)||_2^2 + \frac{\lambda}{m^T\mathbf{1}}\sum_{i = 0}^{K} m_i.\mathcal{L}_{acut}(f_{\theta}(x_i+n_i),x_i) ,
\end{equation}
where $m$ is a masking vector of size K  such that $m_i = 1$ if $x_i$ is a dead leaves image, or $m_i = 0$ otherwise. In order to count the number of dead leaves images we sum this masking vector which is given by $m^T\mathbf{1}$. 

\section{Image denoising results with FFDNet}\label{sec:denoising_acutance}

We choose to train the FFDNet network  \cite{zhang2018ffdnet} to illustrate the impact of the perceptual loss we presented in the previous section.
We adapt the training scheme of the network to the present problem as follows.
First, we increase the size of the training patches from $(50,50,3)$ to $(100,100,3)$.
The reason for this is that the estimation of the 1D-MTF on a small patch is not sufficiently accurate. Keeping the same rings' width would result in fewer estimates for the 1D-MTF. 
On the other hand, decreasing the rings' width would lead to noisier estimates. 
Therefore, we perform the training with larger patches.
Second, we reduce the batch size from 64 to 32 during training to decrease  the memory footprint. We use 150000 samples, made of 100000 natural image patches and 50000 synthesized dead leaves patches.
The other training hyper-parameters remain unchanged, such as the number of epochs or the learning rate decaying schedule.

\subsection{Quantitative evaluation}

In order to show that the proposed scheme indeed has the potential to improve the texture acutance without impairing the usual PSNR evaluation of the performances on natural images, we compute both these metrics for various values of $\lambda$, the weighting parameter in Equation \eqref{eq:lossbatch}. We consider values of $\lambda \in [0,2,5,10,20,50,100,200,500]$. Moreover, we also compute the classical SSIM metric and the perceptual metric PieAPP recently introduced in \cite{prashnani2018pieapp}. 



The models are evaluated numerically on two datasets. 
First, we evaluate the data fidelity by computing the PSNR, SSIM and PieAPP metrics  on the Kodak24 dataset, a benchmark test set of 24 natural images. 
Second, we evaluated the acutance metric on a test set of synthesized dead leaves images.

\begin{figure}[htp]
    \centering
    \includegraphics[width = \textwidth]{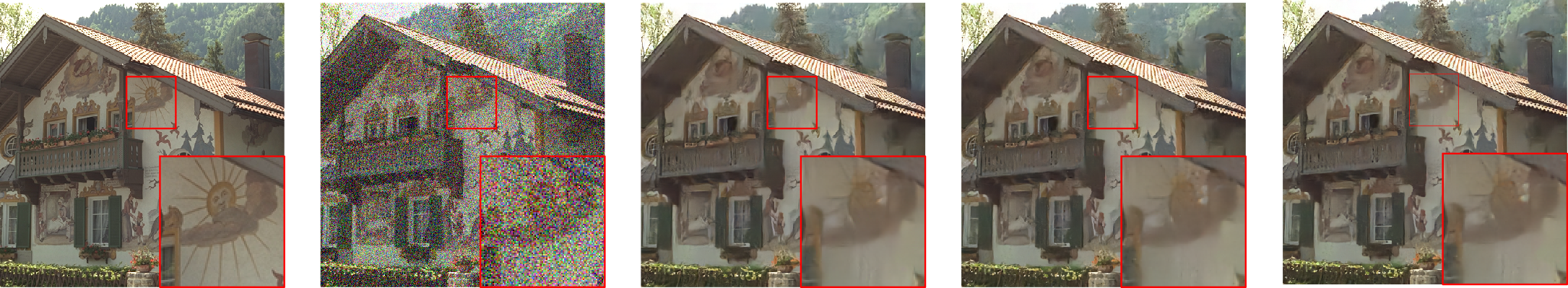}
    
    \includegraphics[width = \textwidth]{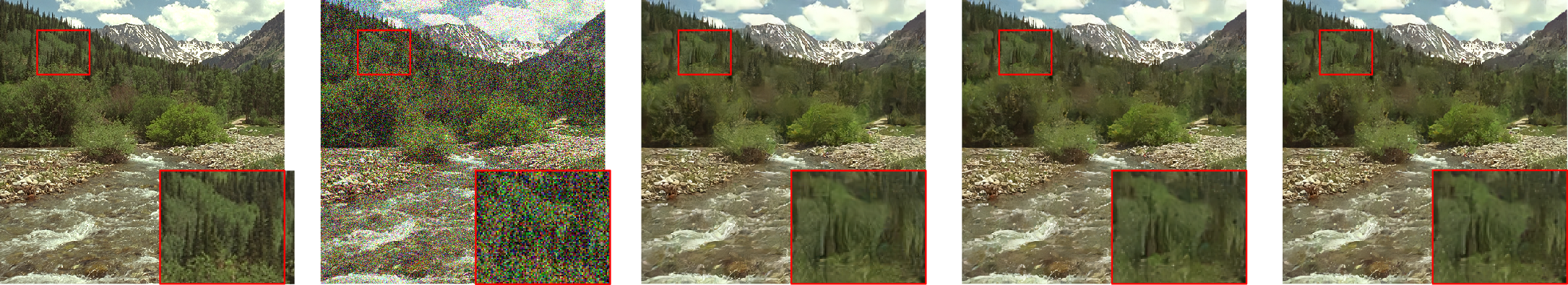}
    \includegraphics[width = \textwidth]{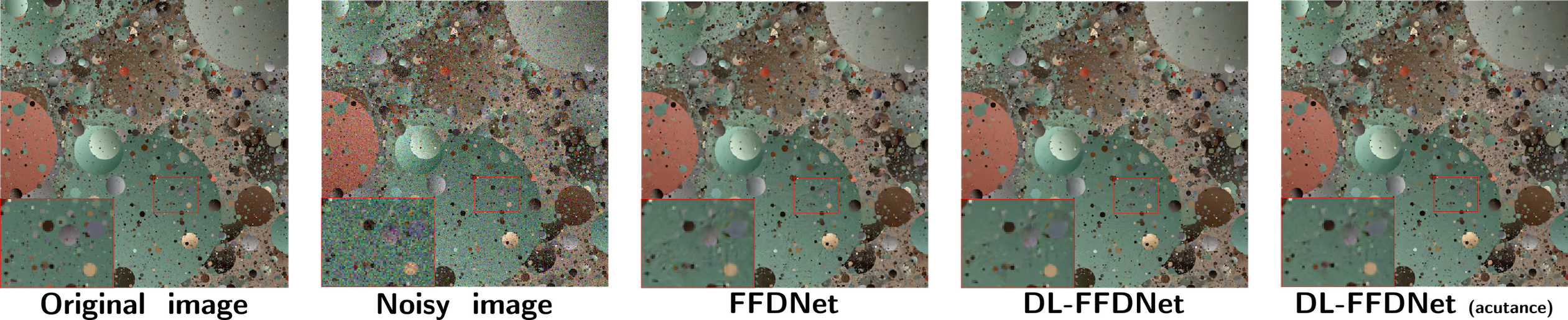}
    \caption[Visual impact of the acutance loss]{ \footnotesize Comparison of FFDNet results on two natural images and on a dead leaves image. From left to right: original image, noisy image, image denoised with standard FFDNet, image denoised with FFDNet trained on a mixed database without the acutance loss, and finally with the acutance loss.}
    \label{fig:mix_vs_nat_comp}
    
\end{figure}

We report, in \cref{tab:table_1}, the numerical evaluation of the trained models. 
We observe a similar behaviour for the tested noise levels $\sigma = 25$ and $\sigma = 50$. 
In both cases, we notice that the standard evaluation metrics, i.e., the PSNR and SSIM, are not affected by the increase of the weighting parameter $\lambda$ until $\lambda = 20$. For values greater than $\lambda = 100$ these metrics decrease rapidly.
On the other hand, the acutance metric keeps improving until $\lambda = 100$ and then reaches a plateau.
This table shows that we can optimize the texture acutance without impairing classic denoising evaluation.
The perceptual evaluation with the PieAPP metric suggests that, for high noise values, the perceptual image quality is slightly enhanced by the addition of the acutance loss.
\begin{table*}[htp]
\centering
\caption{\footnotesize Denoising results of FFDNet trained with different weighting coefficients of the acutance loss for two noise levels. Each cell contains the PSNR, SSIM, PieAPP evaluated on Kodak 24 , and the Acutance metric evaluated on a test set of dead leaves images. Best results in \best{blue}, second results in \second{red}.  }
\label{tab:table_1}
\scriptsize
\begin{tabular*}{\textwidth}{@{\extracolsep{\fill}}| c  r ||c c c c c c c c c |}
\hline
\multicolumn{2}{|c|}{$\lambda$}                                 & 0     & 2    & 5    & 10    & 20   & 50   & 100 & 200 & 500  \\ \hline \hline
\multicolumn{1}{|c|}{\multirow{4}{0.1\textwidth}{$\sigma = 25$}} & PSNR $\uparrow$    & \best{31.88} & 31.87 & \best{31.88} & 31.87 & \best{31.88} & 31.85 & 31.77 & 31.65& 31.56 \\ \cline{2-11} 
\multicolumn{1}{|c|}{}                               & Acutance $\downarrow$  & 0.034 & 0.029 & 0.023 & 0.020 & 0.015 & 0.013& \best{0.012} & \best{0.012}&  \best{0.012}\\ \cline{2-11} 
\multicolumn{1}{|c|}{}                               & SSIM  $\uparrow$    & \best{0.877} &  0.876& 0.875 & 0.875 & \best{0.877} & 0.876& 0.873 & 0.872& 0.869\\ \cline{2-11} 
\multicolumn{1}{|c|}{}                               & PieAPP  $\downarrow$ & \best{0.568} & 0.596 & 0.598 & 0.602 & \second{0.586} & 0.587 &  0.591 & 0.62& 0.612 \\  \hline 
\\
\hline
\multicolumn{1}{|c|}{\multirow{4}{*}{$\sigma = 50$}} & PSNR   $\uparrow$  & \best{28.81} & 28.79 & \second{28.80} & \second{28.80} & \second{28.80} & 28.76 &  28.66 & 28.58& 28.42\\ \cline{2-11}
\multicolumn{1}{|c|}{}                               & Acutance $\downarrow$& 0.084 & 0.078 & 0.073 & 0.053 & 0.035 & 0.026 & \best{0.022} & \best{0.022}& \best{0.022}\\  \cline{2-11}
\multicolumn{1}{|c|}{}                               & SSIM   $\uparrow$  & \best{0.791} & 0.788 & 0.789 & 0.789 & \second{0.790} &  0.789&  0.786&  0.783& 0.779 \\ \cline{2-11} 
\multicolumn{1}{|c|}{}                               & PieAPP  $\downarrow$ & 0.932 & 0.956 & 0.952 & 0.948 & \best{0.912} &  \second{0.925}& 0.934 &  0.940& 0.953\\  \hline
\end{tabular*}
\end{table*}
\vspace{-0.4cm}
Some results can be visualized in \cref{fig:mix_vs_nat_comp} (please zoom in the electronic version of this document). The result with and without using the acutance loss appear quite close, despite the strong improvement of the texture acutance measurement.  
Nonetheless, we can notice some improvements in the preservation of low-contrast details in the first row. Moreover, the contrast is also rendered better when training with the acutance loss.
Finally, on the third row, details on the dead leaves images are better preserved using the acutance loss. On the second and third row, we see that the network trained with natural images sometimes hallucinates details, which are removed when training with dead leaves images.
\subsection{Spectral preservation}

For mixed trainings of FFDNet, the texture acutance score is greatly improved when using the corresponding loss, which is expected. 
However, the acutance score only gives a partial information about the MTF of the trained network. 
In order to further understand the impact of the acutance loss on the spectral preservation ability of the network, we compute its MTF as described next. We compute the 1D-MTF from the denoised image and the original image for each dead leaves image of the synthetic test set. 

 \begin{wrapfigure}{r}{0.5\textwidth}
  \begin{center}
    \includegraphics[width=0.5\textwidth]{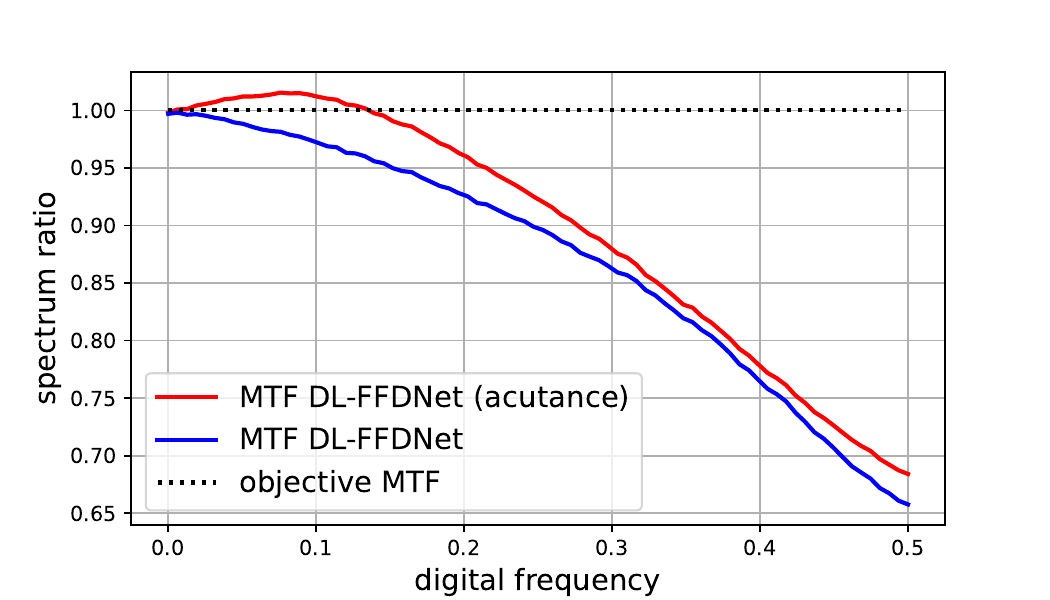}
  \end{center}
\caption[Frequential impact of the acutance loss]{\footnotesize Comparison of the MTF evaluated with FFDNet trained on a mixed database with or without the acutance loss, on the whole dead leaves image test set.}
\label{fig:average_mtf}
\end{wrapfigure} 
Since the 1D-MTF depends on the image's content, which differs from image to image, we average the obtained MTF over the whole dataset.
In \cref{fig:average_mtf}, we report the MTF of FFDNet trained with and without the acutance loss (with $\lambda = 50$) for a noise level $\sigma = 25$. 
Recall that a perfect MTF should be equal to one. We can observe that for low to medium frequency, the MTF of the model trained with the accutance loss is much closer to one.
Actually, the values for low frequency exceed one which is one way the system can improve the acutance and which indeed is a limitation of the approach.
 For high frequency, the gap between the two MTF is smaller, probably as a result of the profile of the CSF function, which quickly decreases for high frequency, see \cref{fig:dead_leaves_spectrum}.
This behavior, as well as the addition of low frequency, could be modified by considering alternative CSF functions and can be easily  integrated into our framework. This could also further improve the preservation of details on examples such as those of \cref{fig:mix_vs_nat_comp}. In this paper, we have decided to keep the original definition of the acutance, since our main goal is to show that this standard measure of the ability to preserve texture can be greatly improved without impairing the other aspects of image quality.

\subsection{RAW image denoising}
As a proof of concept, we extended our experiments to real-world image denoising on the SIDD benchmark \cite{abdelhamed2018high} for cameraphones denoising. To that end, we trained the same denoising network with a U-Net architecture to denoise real RAW images and synthesized RAW dead leaves images. This network produces a RAW denoised image from a RAW input noisy image. To simulate RAW noise for dead leaves images, we used a Poisson-Gaussian model with realistic noise parameters.
Unlike Gaussian noise removal, the loss is here a combination of the $\mathcal{L}_1$ loss and the acutance loss : $\mathcal{L} = \mathcal{L}_1 + \lambda \mathcal{L}_{acut}$. 
For RAW images, the acutance computation differs slightly. 
In order to convert a RAW image to a grey-scale image, we first pack the $(H,W)$ image in a $(H/2,W/2,4)$ RGGB tensor, then we average them in a single $(H/2,W/2)$ grey array, by weighting each channel with the white balance parameters.   
We ran the training for $\lambda \in [0,10,100]$. 
We report the numerical results obtained in \cref{tab:table_2}. 
In comparison with $\lambda = 0$, the PSNR is still good for $\lambda = 10$, while the RAW acutance is largely improved. This improvement also translates in a better acutance in the RGB domain, which was not seen during training. This metric is computed on the denoised images developed with a standard ISP. This experiment shows that we can improve camera evaluation without impairing the image quality in the case of a full camera development pipeline. For $\lambda = 100$, the PSNR noticeably decreases while the RAW acutance reaches a plateau. 

\begin{table}[htp]
\centering
\caption{\footnotesize Denoising results obtained by training a denoising Unet for real RAW images evaluated on the SIDD test set of cameraphone images\cite{abdelhamed2018high}. We report the PSNR, Acutance RAW and acutance RGB metrics. Best results in \textbf{bold.}}
\label{tab:table_2}
\scriptsize
\begin{tabular*}{0.6\textwidth}{@{\extracolsep{\fill}}| c ||c c c  |}
\hline
$\lambda $                                & 0     & 10    & 100    \\ \hline \hline
 PSNR RAW     & \textbf{51.31} & 51.24 & 50.61\\  \hline
 Acutance RAW     &  0.018& \textbf{0.011} &  0.012\\ \hline
 Acutance RGB     & 0.086 &  0.061& \textbf{0.049} \\ 
 \hline
\end{tabular*}
\end{table}

\vspace{-1cm}
\section{Conclusion\label{sec:acutance_conclusion}}

In this work, we have shown that a specific training of image restoration neural networks can greatly improve a standard evaluation metric quantifying the preservation of textures, without impairing classical performance evaluation criteria. As a proof of concept, we extended the use of the acutance loss for real-world image denoising networks, showing that the proposed framework can improve a complete RAW images development pipeline. Considering that the texture acutance metric is routinely used to evaluate digital camera, this founding has potential important practical applications. 

\paragraph{Aknowledgements} This work is supported by the project MISTIC (ANR-19-CE40-005).

\bibliographystyle{splncs04}
\bibliography{bibliography}

\end{document}